\documentclass[lettersize,journal]{IEEEtran}
\usepackage{amsmath,amsfonts}
\usepackage{algorithmic}
\usepackage{algorithm}
\usepackage[skins,listings,raster]{tcolorbox}

\usepackage{array}
\usepackage[caption=false,font=footnotesize,labelfont=rm,textfont=rm,subrefformat=parens,labelformat=parens]{subfig}
 
\usepackage{textcomp}
\usepackage{stfloats}
\usepackage{url}
\usepackage{verbatim}
\usepackage{cite}
\usepackage{multirow}
\usepackage{hyperref}
\usepackage{graphicx} 
\usepackage{booktabs} 
\usepackage{hyperref}
\usepackage{soul}
\usepackage{xcolor}

\usepackage[caption=false,font=footnotesize]{subfig} 
\usepackage{makecell} 
\usepackage{booktabs}

\usepackage{siunitx}
\usepackage[acronym]{glossaries}
\newacronym{mmwave}{mmWave}{millimeter-Wave}
\newacronym{mpnn}{MPNN}{Message Passing Neural Network}
\newacronym{rf}{RF}{Radio Frequency}
\newacronym{rfid}{RFID}{Radio Frequency Identification}
\newacronym{lidar}{LiDAR}{Light Detection and Ranging}
\newacronym{csi}{CSI}{Channel State Information}
\newacronym{rssi}{RSSI}{Received Signal Strength Indicator}
\newacronym{fmcw}{FMCW}{Frequency-Modulated Continuous Wave}
\newacronym{asl}{ASL}{American Sign Language}
\newacronym{drai}{DRAI}{Dynamic Range Angle Images}
\newacronym{msc}{MSC}{Multi-Signal Consolidation}
\newacronym{iq}{IQ}{In-phase and Quadrature}
\newacronym{cnn}{CNN}{Convolutional Neural Network}
\newacronym{lstm}{LSTM}{Long Short-Term Memory}
\newacronym{knn}{KNN}{K Nearest Neighbor}
\newacronym{mlp}{MLP}{Multi-Layer Perceptron}
\newacronym{nll}{NLL}{Negative Log-Likelihood}
\newacronym{cd}{CD}{Chamfer Distance}
\newacronym{ahc}{AHC}{Agglomerative Hierarchical Clustering}
\newacronym{ae}{AE}{Auto-Encoder}
\newacronym{auc}{AUC}{Area Under the Curve}
\newacronym{roc}{ROC}{Receiver Operating Characteristics}

\begin{document}

\title{ImmCOGNITO: Identity Obfuscation in Millimeter-Wave Radar-Based Gesture Recognition for IoT Environments}

\author{Ying Liu, Si Zuo, Chao Yang, Yuqing Song, Dariush Salami, Stephan Sigg,~\IEEEmembership{Senior Member, IEEE}
\thanks{Ying Liu, Si Zuo, Yuqing Song, Dariush Salami, and Stephan Sigg are with the Department of Information and Communications Engineering, Aalto University, 02150 Espoo, Finland. Chao Yang is with the Department of Energy and Mechanical Engineering, Aalto University, 02150 Espoo, Finland.}
}




\maketitle

\begin{abstract}
Millimeter-Wave (mmWave) radar enables camera-free gesture recognition for Internet of Things (IoT) interfaces, with robustness to lighting variations and partial occlusions. However, recent studies reveal that its data can inadvertently encode biometric signatures, raising critical privacy challenges for IoT applications.
In particular, we demonstrate that mmWave radar point cloud data can leak identity-related information in the absence of explicit identity labels.
To address this risk, we propose {ImmCOGNITO}, a graph-based autoencoder that transforms radar gesture point clouds to preserve gesture-relevant structure while suppressing identity cues. The encoder first constructs a directed graph for each sequence using Temporal Graph KNN.
Edges are defined to capture inter-frame temporal dynamics.
A message-passing neural network with multi-head self-attention then aggregates local and global spatio-temporal context, and the global max-pooled feature is concatenated with the original features.
The decoder then reconstructs a minimally perturbed point cloud that retains gesture discriminative attributes while achieving de-identification. Training jointly optimizes reconstruction, gesture-preservation, and de-identification objectives. Evaluations on two public datasets, PantoRad and MHomeGes, show that ImmCOGNITO substantially reduces identification accuracy while maintaining high gesture recognition performance.

\end{abstract}
\glsresetall

\section{Introduction}
{\IEEEPARstart{R}{adio} Frequency (RF)-based gesture recognition has gained increasing attention in the Internet of Things (IoT) domain due to its potential for privacy preservation, unobtrusiveness, and robust sensing capabilities. Unlike traditional optical modalities, RF sensing infers motion from reflections of radio-frequency waves. It remains effective under varied illumination and in the presence of common occlusions such as clothing, furniture, or walls~\cite{xie2024wall}. 
These characteristics make RF-based recognition particularly suitable for IoT environments where vision-based approaches are impractical or raise privacy concerns. 
Its high sensitivity enables IoT systems to capture a broad spectrum of gestures, from coarse hand movements to fine motor actions. 
In healthcare IoT~\cite{ahmed2023machine}, for example, where privacy and hygiene are critical, RF-based gesture recognition supports touch-free patient monitoring and precise motor function tracking in sterile clinical settings. Similarly, in connected vehicles~\cite{kern2023radar}, in-cabin RF monitoring enables intuitive gesture-based control of infotainment systems and enhances passenger safety by detecting unattended children or abnormal driver behavior~\cite{raja2018wibot}. 
Beyond healthcare and automotive, RF-based gesture recognition can play a pivotal role in smart homes, human–computer interaction, and assistive IoT technologies, paving the way for more natural and trustworthy human–IoT interaction.}

\begin{figure}[t]
    \centering
    \includegraphics[width=0.5\textwidth]{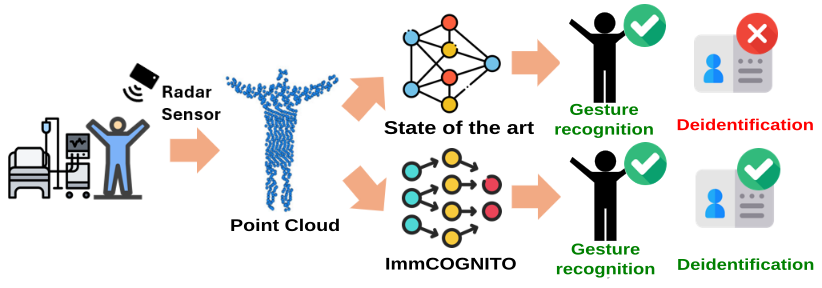}
    \caption{{Radar sensor-based gesture recognition in a privacy-preserving IoT environment. Our approach obfuscates identity while preserving gesture recognition accuracy, enhancing privacy protection.}}
    \label{fig:Fig1}
\end{figure}

{Among these RF-based sensing technologies, \gls{mmwave} radar stands out due to its high operating frequency and correspondingly high spatial resolution. This capability supports reliable detection of both rapid motions and fine-grained gestures.}
These capabilities typically surpass those of lower-frequency RF technologies, such as \gls{rfid} or Wi-Fi~\cite{ling2024uranus}.
Its ability to penetrate non-metallic obstacles further enables gesture recognition in non-line-of-sight scenarios~\cite{pantomime}.
However, this increased sensitivity may inadvertently expose biometric traits, such as vocal cord vibrations~\cite{li2020vocalprint,dong2020secure}, individual breathing patterns~\cite{wang2022your}, and unique gait signatures~\cite{yang2020mu,meng2020gait,chowdhury2023unobtrusive}. 

{{Consequently, radar signals implicitly capture user-specific features alongside the intended sensing data, enabling potential unauthorized identification. This privacy risk raises concerns about personal information leakage and may discourage the adoption of radar-based sensing. Consider a camera-free smart-home gesture interface using a mmWave radar. In real deployments, radar data are often cached or uploaded for debugging and model updates. Even if the system only targets gesture recognition, such traces can be repurposed to infer who performed each interaction, enabling persistent user tracking.
Over time, an unauthorized party can build longitudinal behavioral profiles (e.g., routines, presence patterns, and fine-grained activity habits) and further infer sensitive attributes.}

To validate this threat, we investigate whether public mmWave gesture datasets leak identity information.}
{We demonstrate this risk by repurposing several state-of-the-art gesture recognition models for identification, which include Tesla~\cite{salami2022tesla}, Pantomime~\cite{pantomime}, and PointNet++~\cite{qi2017pointnet++}. 
Specifically, we replace their gesture-classification heads with identity-classification heads and retrain the models using identity labels.}
Comprehensive experiments conducted on the PantoRad and MHomeGes datasets demonstrate that these models can achieve high identification accuracies. 
For example, PointNet++ attains at least 80\% identification accuracy on both datasets. 
{{These findings reveal a critical, previously overlooked problem regarding identity leakage in mmWave radar-based gesture recognition systems. This motivates the need for a method that removes identity features while maintaining the data's utility for the primary sensing tasks.}}

To address the risk, we propose ImmCOGNITO (\textsc{I\textcolor{gray}{dentity obfuscation in}
M\textcolor{gray}{illi}M\textcolor{gray}{eter-wave radar} \textcolor{gray}{gesture re}COGNIT\textcolor{gray}{I}O\textcolor{gray}{N} \textcolor{gray}{for IoT Environments}}),
which transforms point cloud data into a representation that retains gesture attributes while obfuscating identity cues (cf. Fig.~\ref{fig:Fig1}).
ImmCOGNITO leverages a graph-based autoencoder architecture, integrating message passing neural networks and multi-head self-attention mechanisms to model both spatial and temporal dependencies in the point cloud.
The framework is optimized by jointly minimizing gesture recognition loss and point cloud reconstruction loss, while maximizing a de-identification loss that actively suppresses identity-related information.

We comprehensively evaluate ImmCOGNITO on two publicly available mmWave point cloud datasets. The results show that our approach can effectively reduce person identification accuracy while maintaining high gesture recognition performance.
Focusing on the dual objectives of gesture recognition and identity obfuscation, ImmCOGNITO applies minimal yet targeted modifications to the original point clouds. These changes thereby preserve the utility of the data across various downstream tasks beyond gesture recognition.
In summary, the main contributions of this work are as follows:
\begin{itemize}
\item {We demonstrate that publicly available \gls{mmwave} point cloud datasets curated for gesture recognition can be repurposed for person identification. This is accomplished by fine-tuning gesture recognition models to extract person-specific execution patterns.}

\item We introduce a graph-based autoencoder that leverages message passing neural networks and multi-head self-attention to learn identity-obfuscated representations while maintaining gesture recognition performance.

\item{We conduct empirical evaluations which show that ImmCOGNITO significantly reduces identification accuracy without compromising gesture recognition. These results suggest a path toward de-identification in radar sensing.}

\end{itemize}


\section{Related Work} \label{section2}


Vision-based sensing technologies (e.g., RGB and depth cameras) recognize human gestures from spatial and temporal features in the visual sequence~\cite{visionsurvey2}. 
Liao et al. proposed to detect hand contours and motion trajectories to improve gesture tracking accuracy~\cite{7379635}.
Zhu et al. extended this by incorporating adaptive filtering and motion analysis, enhancing robustness against background noise and lighting variations~\cite{Zhu2013Vision}.
Liu et al. further refined gesture recognition by leveraging deep convolutional and recurrent architectures, which improve accuracy and generalization across diverse persons and motion patterns~\cite{9134375}.
Shahi et al. introduced a learning-based framework that enabled person-customizable gestures from a single demonstration, reducing the need for extensive training data while improving adaptability~\cite{shahi2024vision}.
Despite these achievements, vision-based gesture recognition faces fundamental challenges in low-light conditions and occluded scenarios~\cite{chakraborty2018review}, and its reliance on cameras raises privacy concerns~\cite{zhao2023visual}.

Radio Frequency (RF)-based sensing (e.g. \gls{rfid} and Wi-Fi) is an alternative that addresses some of the privacy concerns.
Recently, Zhang et al. proposed an accurate real-time gesture recognition system for \gls{rfid}, which is robust to changes in gesture execution speed~\cite{rfid1}.
The authors exploited time and frequency domain features to segment and detect 18 hand gestures in close proximity to a grid of four \gls{rfid} tags.
Ma et al. added siamese networks to significantly reduce the number of gesture samples needed for training~\cite{rfid2}.
Moreover, Dian et al. achieved greater gesture diversity and domain independence through multimodal convolutional neural networks and adversarial learning~\cite{rfid3}.
Bu et al. further demonstrated the distinction of touch gestures on a surface in a battery-free interaction system using a grid of passive \gls{rfid} tags~\cite{rfid4}.
Despite these achievements, \gls{rfid} based sensing systems are constrained by their short operational range~\cite{zhao2022wear} and are sensitive to tag placement~\cite{liu2021rfid}.

Wi-Fi-based gesture recognition has garnered significant attention for non-intrusive and accurate sensing.
Gu et al. proposed a dual-attention network to enhance feature extraction by isolating relevant signal regions for improved recognition accuracy even in complex environments~\cite{wifi2}.
Gao et al. presented a meta-motion-based framework for continuous gesture recognition, enabling the recognition of gesture streams without explicit segmentation or pauses~\cite{gao2024wicgesture}.
However, Wi-Fi-based gesture recognition is challenged by environmental interference and multipath effects, which impair the accuracy and reliability of complex gesture detection~\cite{wifi5}. The limitations of vision, \gls{rfid}, and Wi-Fi-based systems can be addressed by \gls{mmwave} radar sensing, particularly \gls{fmcw}~\cite{gupta2021target}.

MmWave radar sensing has demonstrated high accuracy and robustness of gesture recognition~\cite{liu2024pmtrack}.
For instance, Santalingam et al. focused on employing spatial spectrograms with \qty{60}{\giga\hertz} \gls{mmwave} signals for robust \gls{asl} recognition, achieving \qty{87}{\percent} accuracy~\cite{mmASL}.
Li et al. employed spatio-temporal processing to capture signal variations from human gestures while integrating data augmentation and real-time segmentation~\cite{mmwave2}.
Hayashi et al. optimized gesture recognition exploiting multiple radar swipes to achieve temporal gesture information~\cite{radarnet}.
The authors in~\cite{mgesture} achieved person-independent recognition with less training data by isolating gesture-specific features and suppression of personalized discrepancy.
In general, \gls{mmwave} radar systems are able to capture micrometer-level vibrations~\cite{IQradar}.
Wang et al. proposed a time-distributed CNN-transformer network (DCS-CTN) that leverages the raw mmWave radar data cube, including phase information, to achieve high-precision subtle gesture recognition, reporting an accuracy of 99.75\%\cite{DCSCTN}.
The above studies employ \gls{iq} data, range Doppler maps, and \gls{drai} and typically result in high data volumes that need to be processed for gesture recognition systems.
For real-time processing and especially on embedded devices, the implementation of these data-intensive methods is therefore challenging and often impossible.

On the other hand, point cloud-based methods are more efficient and hence allow larger datasets for training.
For example, the authors of~\cite{pantomime} exploited spatio-temporal properties of \gls{mmwave} RF signals with sparse 3D point clouds for mid-air gesture recognition, achieving high accuracy and real-time performance with low computational demands.
Similarly, Salami et al. presented a gesture recognition system that utilizes message passing neural networks for sparse point clouds from \gls{mmwave} radars, which significantly reduces the computational complexity while maintaining high accuracy~\cite{salami2022tesla}.
Liu et al. leveraged transfer learning to process point clouds for gesture recognition across different environments and individuals~\cite{mTransSee2021}.
Additionally, authors in~\cite{radhar} utilized sparse and non-uniform point clouds from \gls{mmwave} radars to achieve high accuracy in identifying human activities.

MmWave radar systems are capable of capturing {biometric information.}
For instance, the VocalPrint system leveraged mmWave radar to capture vocal vibrations from the throat area, isolating these signals from ambient noise using a resilience-aware clutter suppression technique, thereby enabling robust authentication~\cite{li2020vocalprint}.
Dong et al. combined radar-detected vocal cord vibrations and lip motion to create a multimodal biometric system that enhances security for Internet of Things smart homes by ensuring liveness detection and resisting spoofing attacks~\cite{dong2020secure}.
Wang et al. captured unique breathing patterns, employing a rotating gadget to optimize signal capture from multiple individuals, thereby enhancing multi-person authentication~\cite{wang2022your}.
Similarly, Yang et al. utilized the radar's high directivity to identify multiple persons through distinctive gait patterns analyzed in the range-Doppler domain~\cite{yang2020mu}.
Meng et al. further advanced gait recognition by collecting extensive mmWave data on gait patterns~\cite{meng2020gait}.
Cheng et al. focused on extracting spatio-temporal information from 4-D radar point cloud data for person reidentification (ReID)~\cite{cheng2021person}.
Furthermore, Zhao et al. implemented a radar-based human tracking and identification system which utilizes sparse point clouds and deep recurrent networks to achieve high tracking and identification accuracy~\cite{zhao2019mid}.
Chowdhury et al. captured unique gait patterns through point clouds augmented with novel height surface maps to enhance identification accuracy in closed spaces using a combination of PointNet architectures and majority voting schemes~\cite{chowdhury2023unobtrusive}.
The effectiveness of person identification in \gls{mmwave} based systems raises privacy concerns and motivates de-identification to preserve data usability while preventing unauthorized identification.

De-identification in image-based systems has explored techniques such as blurring~\cite{maity2023preserving}, pixelation~\cite{cao2024face}, and transformation-based methods~\cite{bae2022selective}
to obscure identities while preserving data utility.
Studies addressing these challenges in radar-based systems are yet limited. Liu et al. proposed mmFilter, a local perturbation module that selectively disrupts radar data to disable posture estimation while preserving gesture recognition~\cite{liu2023application}. However, its permutation-based approach degrades gesture recognition accuracy and is limited to hiding posture information, making it less effective for identity obfuscation. 
Our work aims to specifically protect identity within gesture recognition systems by obfuscating identity information without compromising gesture recognition accuracy.

\section{Problem Formulation}\label{section3}
We aim to transform mmWave point cloud data to retain gesture information while removing privacy-related information, such as identity \footnote{note that the described process is applicable also for arbitrary other privacy-related information, not only for identity as demonstrated here.}.
Given a point cloud $p$, let $u$ represent the identity attributes to be obfuscated and $g$ represent the gesture attributes to be preserved. The overall objective is to define a function $\Phi$ that applies perturbations to the input point cloud $p$, producing an output point cloud $p' = \Phi(p)$, such that $p'$ satisfies the following properties:
\begin{itemize}
    \item \textbf{Preservation of Gesture Attributes:} 
    The gesture attributes \( g \) should be effectively preserved in the perturbed point cloud \( p' \).
    In particular, the gesture recognition classifier \( G \) is expected to maintain high performance when operating on the perturbed point cloud \( p' \).

    \item \textbf{Obfuscation of Identity:}
    The identity attributes \( u \) should be obfuscated to significantly degrade the performance of the identification classifier \( U \) when applied to the perturbed point cloud \( p' \).
    This implies that after applying the function \( \Phi \), the generated point cloud \( p' \) should substantially diminish the accuracy and reliability of the classifier \( U \) designed for identifying persons based on their unique identity attributes.

    \item \textbf{Retention of Point Cloud Structure}:
    The structure of the perturbed point cloud $p'$ and the coordinates within it do not change significantly compared to the original point cloud $p$.
    The function $\Phi$ should introduce minimal alterations to the overall geometric structure of the point cloud.
    This constraint is intended to preserve data versatility and support applicability across diverse downstream tasks and frameworks.
    
\end{itemize}

\begin{figure}
  \centering
  \includegraphics[width=\linewidth]{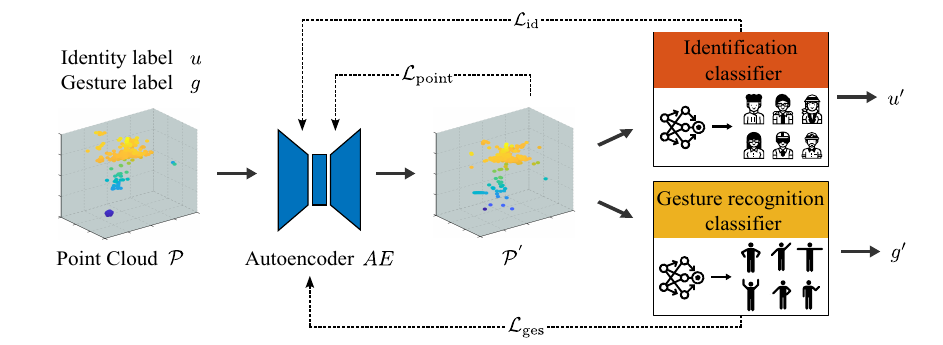}
  \caption{Architecture of ImmCOGNITO for point cloud processing with de-identification and gesture feature preservation.}
  \label{fig:ae_framework}
\end{figure}

\section{Methodology} \label{section4}

Fig.~\ref{fig:ae_framework} illustrates the workflow of the framework.
First, point cloud data \(p\), which includes gesture labels \(g\) and identity labels \(u\), is processed by the Autoencoder, which aims to generate an output point cloud \(p'\).
The primary objective of the Autoencoder is to preserve gesture features while obfuscating identity information in \(p'\).
The output point cloud \(p'\) is evaluated using two pre-trained classifiers: a gesture recognition classifier \(G\) and an identification classifier \(U\).
Each classifier predicts corresponding labels -- \(G\) predicts gesture labels \(g'\), and \(U\) predicts identity labels \(u'\) -- by generating probability distributions from \(p'\).
These distributions are used to compute the gesture recognition loss \(\mathcal{L}_{\text{ges}}\) and the identification loss \(\mathcal{L}_{\text{id}}\), respectively.
While \(\mathcal{L}_{\text{ges}}\) ensures that the Autoencoder retains essential gesture features, \(\mathcal{L}_{\text{id}}\) promotes misclassification of identities, guiding the Autoencoder to eliminate identity-specific features.
Consequently, the Autoencoder ensures that the predicted gesture labels \(g'\) closely align with the actual labels \(g\).
Simultaneously, it modifies the identity signals in the point clouds, ensuring that the predicted identity labels \(u'\) diverge from the actual labels \(u\).

In addition, the Autoencoder minimizes the point cloud reconstruction loss \(\mathcal{L}_{\text{point}}\) to encourage that the output point cloud \(p'\) maintains spatial and geometric consistency with the input point cloud. This preservation aims to better preserve its fundamental structure, thereby maintaining the potential for its utility in various applications beyond gesture recognition.


In the training process of the Autoencoder, the weights of both the gesture recognition classifier \(G\) and the identification classifier \(U\) are fixed.
Consequently, variations in the gesture recognition loss \(\mathcal{L}_{\text{ges}}\) and the identification loss \(\mathcal{L}_{\text{id}}\) are attributed solely to the Autoencoder's performance modifications.


\begin{figure*}[t]
  \centering
  \includegraphics[width=\textwidth]{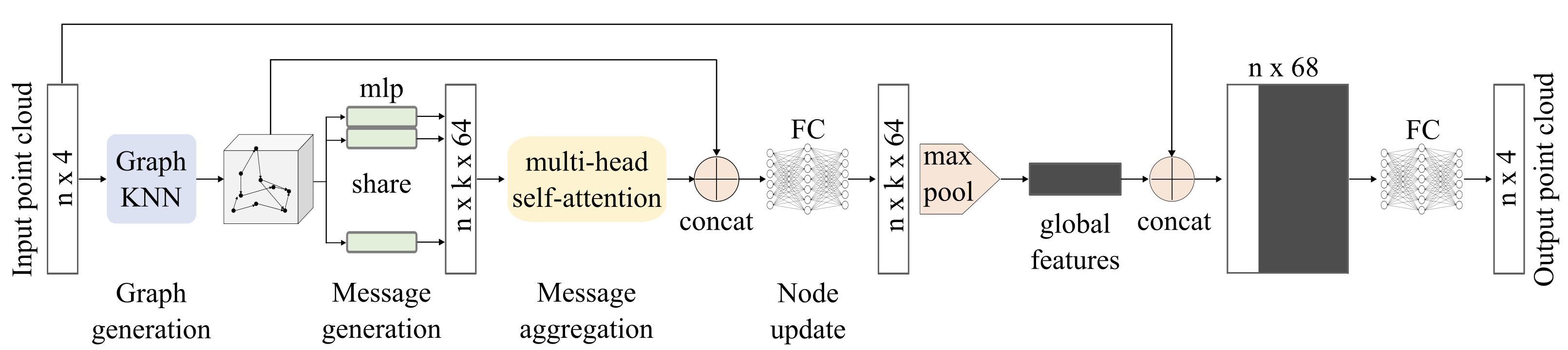}
  \caption{Autoencoder architecture. Starting with $n$ points as input, the network uses Temporal Graph \gls{knn} to transform it into a directed graph. The graph is processed through a message passing neural networks layer with three phases: message generation (using \glspl{mlp}), message aggregation (with multi-head self-attention), and node update (via fully connected layers). Global features are extracted via max pooling, concatenated with original features, and output a point cloud with $n$ points . "mlp" denotes multi-layer perceptron, "concat" denotes concatenation operation, and "FC" represents fully connected layers with ReLU activation.}
  \label{fig:fig3}
\end{figure*}

\subsection{Autoencoder Module}
Building upon semi-adversarial methods~\cite{mirjalili2020privacynet}, we propose an Autoencoder framework specifically designed to modify point clouds while preserving essential spatial and temporal features critical for gesture recognition.
Reconstructing point clouds, which encode gestures, is crucial as each point cloud comprises a sequence of frames, with each frame consisting of an unordered set of points in a 3D space.
Unlike prior work that primarily focused on reconstructing static images~\cite{liu2023enhanced} or point clouds~\cite{yang2019pointflow} by leveraging spatial properties, our approach tackles both the spatial and temporal dependencies embedded in dynamic point cloud sequences.
This challenge arises from the temporal correlations and subtle spatial-temporal patterns within each frame, which are critical for accurate gesture recognition. Consequently, any perturbations introduced during reconstruction must maintain temporal coherence and preserve the intricate spatial-temporal characteristics of the input.

To effectively model the spatial-temporal dependencies inherent in point clouds, we propose a graph-based Autoencoder, depicted in Fig.~\ref{fig:fig3}.
We first form a directed graph from the input point cloud data, where each edge specifically encodes temporal relationships between points across sequential frames.
Utilizing message passing neural networks in conjunction with a multi-head self-attention mechanism, the Autoencoder extracts and processes high-dimensional features that capture both spatial and temporal aspects.
These features are then concatenated with the original features.
This reconstructed point cloud preserves the integrity of spatial and temporal data by retaining essential gesture attributes, while achieving de-identification.

Assume a gesture based point cloud \( \mathcal{P} = \{p_i \mid p_i = (x_i, y_i, z_i, t_i), \, i = 1, 2, \ldots, N\}\), in which \( (x_i, y_i, z_i) \) represent \num{3}D coordinates and \( t_i \in \{1, 2, \dots, F\} \) is the frame index that indicates the temporal frame to which the point $p_i$ belongs.
We utilize the Temporal Graph \gls{knn} algorithm~\cite{salami2022tesla} to transform the point cloud into a directed graph. {{This graph construction is dynamic and data-driven, meaning the topology is determined by the specific spatiotemporal distribution of each input point cloud rather than a predefined structure.}}
This transformation retains the original features of the point cloud while simplifying further processing.
We define the directed graph as $\mathcal{T} = (p, E)$, where \( p \) represents the nodes corresponding to all points in the point cloud, and \( E \) constitutes the set of directed edges.

Directed edges reflect temporal features across significant frames. For each point \( p_i \), edges are drawn from its \( k \) nearest neighbors in the preceding frame, based on distance metrics.
When $p_i$ belongs to the first frame, i.e., $t_i = 1$, the edges are instead drawn to neighbors within the current frame.
The potential neighbor set is expressed as
\begin{equation}
    C(p_i) = 
\begin{cases} 
\{ p_j \mid p_j \in \mathcal{P}, t_j = t_i - 1 \}, & \text{if } t_i > 1, \\
\{ p_j \mid p_j \in \mathcal{P}, t_j = t_i \}, & \text{if } t_i = 1.
\end{cases}
\end{equation}

We utilize Euclidean distance to quantify the distance between two points within the point cloud.
Specifically, the distance \( d_{ij} \) between point \( p_i \) and each point \( p_j \) within the potential neighbor set \( C(p_i) \) is calculated based on their \num{3}D coordinates (the Euclidean distance between points of adjacent frames when collapsing the time difference between the frames):

\begin{equation}
d_{ij} = 
        \sqrt{(x_j - x_i)^2 + (y_j - y_i)^2 + (z_j - z_i)^2}, p_j \in C(p_i).
\end{equation}

Upon calculating those distances, the $k$ points in $C(p_i)$ with the smallest distances to $p_i$ are selected as the neighbors of point $p_i$.
We define the neighbor set as $D(p_i)$.
Then, for each point $p_i$, $k$ directed edges from the points in $D(p_i)$ to point $p_i$ are established. Consequently, the edge set \(E\) of the graph comprises up to \(k \times N\) edges, where \(N\) is the number of points in the point cloud.
This edge set is formally expressed as \(E = \{(p_j, p_i) \mid p_j \in D(p_i)\}\),  where $(p_j, p_i)$ represents the directed edge from the neighboring point \(p_j\) to point \(p_i\).
This graph construction is intended to encode the temporal dimension of the data by emphasizing continuity and the evolution of features across frames.
{{It is noted that while the graph is dynamically generated for each sample to encode temporal continuity, the edge connections are deterministic based on geometric proximity and remain fixed during the network training process (i.e., the adjacency matrix is not a learnable parameter).}}

Following the construction of the graph \( \mathcal{T} = (p, E) \), the spatial and temporal features of the data are analyzed using message passing neural networks~\cite{mpnn_original}.
In message passing neural networks, information is processed through three principal stages: message passing, message aggregation, and node updating.

Prior to employing message passing neural networks, the initial node features are derived from the coordinates of each point and encoded using a multi-layer perceptron (MLP) $f_{\text{mlp}}$, defined as:
\begin{equation}
    h_i = f_{\text{mlp}}(x_i, y_i, z_i).
\end{equation}

We chose an \gls{mlp} to encode node features to capture complex non-linear relationships within the data, which enhances the initial embeddings and facilitates a deeper understanding of the graph's intricate structure.

In the message generation phase, each directed edge \( (p_j, p_i) \in E \) from point $p_j$ to $p_i$ is processed using a message generation function \( M \), defined as a fully connected network.
This network is shared across all nodes, employing unified weights to reduce the number of trainable parameters and to enhance the model's generalizability across different node configurations within the graph.
The input for this function is the concatenated vector \( (h_i \oplus (h_j - h_i)) \), which combines the features of node \( p_i \) with the difference of features between \( p_i \) and its neighbor \( p_j \). 
Specifically, the message generation for an edge is performed as follows:
\begin{equation} 
m_{ij} = M\left(h_i\oplus (h_j - h_i)\right),
\end{equation}
where \( M \) maps the concatenated features to generate the message \( m_{ij} \), facilitating detailed feature interactions between connected nodes.

In the message aggregation phase, a multi-head self-attention mechanism~\cite{vaswani2017attention} is utilized. The multi-head self-attention mechanism enhances the model's ability to focus on different parts of the input data simultaneously. 
It integrates messages from diverse regions of the graph, which can help capture interdependencies among nodes relevant to structural modeling.

Specifically, for each attention head $b$, the messages $m_{ij}$ are linearly transformed by the corresponding weight matrices to generate sets of queries $Q$, keys $K$, and values $V$. 
For point $p_i$, the queries $Q_b^{(i)}$, keys $K_b^{(i)}$, and values $V_b^{(i)}$ are computed as follows: 
\begin{equation}
Q_b^{(i)} = W_b^Q m_{ij}, \quad K_b^{(i)} = W_b^K m_{ij}, \quad V_b^{(i)} = W_b^V m_{ij}, 
\end{equation}
where $W_b^Q, W_b^K, W_b^V$ are the weight matrices for the $b$-th head. 
Attention score $\alpha_{ij}$ between node $p_i$ and its neighbor $p_j$ is then calculated by: 
\begin{equation} 
\alpha_{ij}^{(b)} = \frac{\exp\left(\frac{Q_b^{(i)} \cdot K_b^{(j)T}}{\sqrt{k}}\right)}{\sum_{j' \in \mathcal{N}(i)} \exp\left(\frac{Q_b^{(i)} \cdot K_b^{(j')T}}{\sqrt{k}}\right)},
\end{equation}
where $\mathcal{N}(i)$ denotes the set of neighboring nodes of $p_i$, and $k$ is the dimension of the queries $Q_b^{(i)}$ and keys $K_b^{(i)}$, reflecting the number of neighbors each node has in the graph $D$.
The attention scores are leveraged to aggregate messages from the respective neighbors for each attention head $b$: \begin{equation} 
z_i^{(b)} = \sum_{j \in \mathcal{N}(i)} \alpha_{ij}^{(b)} V_b^{(i)}.
\end{equation} 
 This aggregation mechanism effectively captures the weighted significance of each neighbor’s features. Subsequently, to synthesize the diverse representational focuses of each head, the vectors $z_i^{(b)}$ from all heads are concatenated and then linearly transformed: \begin{equation} 
 z_i = (z_i^{(1)}\oplus z_i^{(2)}\oplus ... \oplus z_i^{(B)})W^O, 
 \end{equation} 
 where $W^O$ is the output transformation matrix that integrates the multi-headed outputs into a unified feature representation for node $p_i$, with $B$ indicating the number of attention heads.

In the node updating phase, node features are refined using the aggregated messages. The update mechanism employs a fully connected network, denoted as \( f_{\text{update}} \), which processes the concatenated current node features \( h_i \) and the aggregated messages \( z_i \) through linear transformations followed by non-linear activations. The updated node features are defined as:
\begin{equation} 
h_i' = f_{\text{update}}[(h_i \oplus z_i)].
\end{equation}

Following the feature update, a max pooling operation is applied across the entire point set to aggregate the features, enhancing the representation with global contextual information. The result of the max pooling across all nodes is computed as:
\begin{equation}
H_{\text{max}} = \max_{0 \le i < n} \, h_i',
\end{equation}
where $\max$ denotes the max pooling operation.
These aggregated features, represented by \( H_{\text{max}} \), are subsequently concatenated with the initial position feature \( p_i \), forming an enriched feature set. This set undergoes further processing through multiple fully connected layers, designated as \( f_{out} \), to produce the de-identified point cloud \( p' \). The final transformation is represented by:
\begin{equation}
p' = f_{out}\left(p_i \oplus H_{\text{max}}\right).
\end{equation}

\subsection{De-identification}
The primary goal of the de-identification process is to modify the point clouds via the Autoencoder to effectively obscure identity while maintaining the accuracy of gesture recognition. 
To achieve this, a specialized de-identification loss $\mathcal{L}_{\text{id}}$ is employed, with the objective of maximizing the misclassification rate of the identification classifier $U$ on the generated point clouds. To encourage incorrect predictions of identity, the de-identification loss $\mathcal{L}_{\text{id}}$ is defined as the negative of the \gls{nll} loss:
\begin{equation}
\mathcal{L}_{\text{id}} = \frac{1}{M} \sum_{k=1}^{M} \log P(u^{(k)} \mid p^{(k)}),
\end{equation}
where \( M \) is the batch size, representing the number of point clouds in a batch, and \( P(u^{(k)} \mid p^{(k)}) \) denotes the probability that the classifier \( U \) predicts the actual identity class \( u^{(k)} \) for each point cloud \( p^{(k)} \). By maximizing the $\mathcal{L}_{\text{id}}$, we effectively encourage the classifier \( U \) to assign lower probabilities to the correct identity class, thereby enhancing the effect of de-identification.


During training, we observed that the original de-identification loss could exhibit numerical instability. Accordingly, we reformulated it to enforce positivity and improve stability.
To address this, we introduced the logarithmic function $\log(1 + x)$ and an offset $\delta$ to regulate the rate of constraint growth, which are crucial for preventing gradient explosion during the training phase.
The enhanced de-identification constraint function is formulated as follows:
\begin{equation} 
\mathcal{L}'_{\text{id}} = -\log \left(1 + \mathcal{L}_{\text{id}} \right) + \delta ,
\end{equation}
 where $\delta$ is an offset introduced to ensure that the loss function maintains a positive value.
 Note that while $\delta$ shifts the overall value of the loss function, it does not influence the gradient and thus does not affect the direction or magnitude of gradient updates during backpropagation.
 This formulation not only enhances the de-identification effect but also ensures the stability and efficiency of training.
 
\subsection{Gesture Recognition Preservation}
While the de-identification process aims to obscure identity, it is essential to ensure that this modification does not compromise the gesture recognition capability of the generated point clouds.
Therefore, we introduce a gesture recognition loss.
It is implemented by passing the generated point clouds through a gesture recognition classifier.
During Autoencoder training, minimizing the gesture-recognition loss encourages the generated point clouds to retain essential gesture features.
During evaluation, the same classifier is applied to both the original and generated point clouds to quantify the impact of de-identification on gesture recognition accuracy.
In this study, we utilize Tesla~\cite{salami2022tesla} as the gesture-recognition classifier.
Similar to the de-identification loss, we minimize the gesture-recognition loss during training, which is defined as the negative log-likelihood loss:
\begin{equation}
\mathcal{L}_{\text{ges}} = -\frac{1}{M} \sum_{k=1}^{M} \log P(g^{(k)} \mid p^{(k)}),
\end{equation}
where \( M \) is the batch size, and \( P(g^{(k)} \mid p^{(k)}) \) denotes the probability that the gesture recognition classifier \( G \) predicts the actual gesture class \( g^{(k)} \) for each point cloud \( p^{(k)} \).
Minimizing $\mathcal{L}_{\text{ges}}$ guides the Autoencoder to preserve essential gesture-related information in the point clouds while applying de-identification modifications. 

\subsection{Point Cloud Reconstruction}
We also introduce a point cloud reconstruction loss $\mathcal{L}_{\text{point}}$.
This loss encourages the Autoencoder to preserve essential spatio-temporal characteristics of the input point clouds during reconstruction.
We define the point cloud reconstruction loss $\mathcal{L}_{\text{point}}$ using the Chamfer Distance~\cite{chamfer_distance}.
This metric assesses the similarity between two point clouds by calculating the minimum bidirectional point-to-point distances.
$\mathcal{L}_{\text{point}}$ represents the Chamfer Distance between the original point cloud $p$ and the reconstructed point cloud $p'$:
\begin{equation}
\mathcal{L}_{\text{point}} = \sum_{p_m \in p} \min_{p_n \in p'} \|p_m - p_n\|_2^2 + \sum_{p_n \in p'} \min_{p_m \in p} \|p_n - p_m\|_2^2,
\end{equation}
The operation $\| p_n - p_m \|_2^2$ represents the squared Euclidean distance between $p_m$ and $p_n$.
This loss penalizes discrepancies between corresponding point pairs in the original and reconstructed point clouds.
Minimizing $\mathcal{L}_{\text{point}}$ promotes preservation of global structure and geometry during reconstruction.


\subsection{Training Loss}
To achieve a balanced optimization between point cloud reconstruction, gesture recognition preservation, and de-identification, we propose an integrated training loss function. The loss aggregates reconstruction, gesture recognition, and de-identification losses to jointly optimize these objectives:
\begin{equation} 
\mathcal{L}_{\text{final}} = \alpha \cdot \mathcal{L}_{\text{point}} + \beta \cdot \mathcal{L}_{\text{ges}} + \gamma \cdot H(A_{\text{id}}(t) - \tau) \cdot \mathcal{L}'_{\text{id}}.
\end{equation} 
Here, $\alpha$, $\beta$, and $\gamma$ are scaling factors that adjust the relative importance of each loss component in the overall training process, and $A_{\text{id}}(t)$ denotes the identification accuracy at training time $t$.
$H(x)$ is the Heaviside step function, defined as $H(x) = 1$ if $x \geq 0$ and $H(x) = 0$ otherwise.
The step function $H(x)$ is used to conditionally activate the identification loss $\mathcal{L}_{\text{id}}$.
Specifically, $\mathcal{L}_{\text{id}}$ is included in the final loss only when $A_{\text{identity}}(t)$ exceeds the threshold $\tau$. In such cases, $H(x) = 1$.
If $A_{\text{id}}(t)$ is below $\tau$, $H(x) = 0$, thereby deactivating the identification loss.
This selective activation can reduce unnecessary computational overhead, potentially accelerating the training process.

\section{Experiments}\label{section5}
\subsection{Datasets and Evaluation Metrics}
\label{section:dataset}

To validate the effectiveness of ImmCOGNITO, we utilize two publicly available \gls{mmwave} radar-based point cloud datasets.
We briefly summarize each dataset and the evaluation protocol below.
{Pantomime-Radar (PantoRad) dataset~\cite{pantomime}:} The PantoRad dataset contains \num{7208} samples representing \num{21} types of mid-air gestures, performed by \num{41} participants.
Data collection spans various anchor positions from \qtyrange{1}{5}{\meter}, and the majority of data is collected at \qty{1}{\meter}. We partition the \num{21} gestures into three evaluation subsets: \textit{Single-Hand}, \textit{Two-Hand} and \textit{All-Gestures}. 
The \textit{Single-Hand} subset contains \num{9} gestures performed with one hand. 
The \textit{Two-Hand} subset comprises \num{12} gestures that require two-hand coordination or explicit linear or circular motion. 
The \textit{All-Gestures} subset includes all \num{21} gestures. 
{MHomeGes dataset~\cite{MHomeGes}:} Comprising \num{21487} samples, this dataset captures a wide range of arm movements performed by \num{25} volunteers across \num{13} different anchor points, performing \num{10} different categories of gestures with distances ranging from \qtyrange{1.2}{3}{\meter}.

The two datasets cover diverse scenarios, including offices, meeting rooms, homes, and open spaces, and feature variations in subject physiology and gestures. Both datasets were captured using a \qty{77}{\giga\hertz} IWR1443 \gls{mmwave} radar with varying frame rates. A split of \qty{70}{\percent} for training and \qty{30}{\percent} for testing is applied.
We evaluate gesture recognition and de-identification on the original and de-identified point clouds of both datasets using accuracy, $F_1$-score, AUC, and the confusion matrix.
\subsection{Baselines and Evaluation Protocols}
To evaluate the performance of ImmCOGNITO in maintaining gesture recognition accuracy and achieving de-identification, we employ three state-of-the-art radar point cloud-based gesture recognition algorithms. 
{We employ three state-of-the-art mmWave radar point cloud-based backbones as baselines in our evaluation: {Tesla}~\cite{salami2022tesla}, {Pantomime}~\cite{pantomime}, and {PointNet++}~\cite{qi2017pointnet++}. 
{Tesla} is a lightweight gesture recognition framework designed for mmWave point clouds, combining spatial feature extraction with temporal sequence modeling. 
{Pantomime} leverages PointNet++ for spatial feature extraction and Long Short-Term Memory (LSTM) layers to capture temporal dynamics, making it effective for gesture recognition from sparse point cloud sequences. 
{PointNet++} is a widely used point cloud processing model that learns hierarchical local-to-global spatial representations and is particularly strong in capturing complex spatial patterns.}

These algorithms are specifically designed for gesture recognition tasks with high effectiveness.
Consequently, they are selected to evaluate autoencoder de-identified point clouds in the context of gesture recognition. By modifying their final output layers, we adapt them for identification tasks. In Section~\ref{section:UI}, we present their performance on identification across two datasets.
The results show that, although not originally designed for this purpose, these algorithms perform identification effectively.
{For clarity, we append “-GR” to a model name when it is used as a gesture recognition evaluator of both the original and autoencoder-generated point clouds, i.e., {Tesla-GR}, {Pantomime-GR}, and {PointNet++-GR}. We append “-ID” when the same backbone is used for identification, yielding {Tesla-ID}, {Pantomime-ID}, and {PointNet++-ID}. In terms of usage, the -GR variants are employed wherever gesture recognition accuracy is reported, whereas the -ID variants are used in Section~\ref{section:UI} for identification benchmarking across the two datasets and throughout subsequent experiments to quantify de-identification.}


\subsection{Implementation Details}
In this study, point cloud data from the two datasets are processed into sequential frames based on the reception time of each point.
Let $N$ be the total number of points in the point cloud and $F$ as the number of frames desired.
The total number of points \( N \) is first ordered by reception time and divided into segments, with the first ${N}/{F}$ points assigned to the first frame, the next \( {N}/{F} \) points to the second frame, and so forth, maintaining the temporal sequence throughout.
To ensure an equal number of points per frame for each gesture while preserving the original shape of the point cloud, a resampling strategy is applied.
For point reduction, K-means clustering
is used, and cluster centroids are selected as representative points.
In the case that points need to be added, agglomerative hierarchical clustering
is iteratively applied, adding centroids as new points until the desired number of points is achieved.
In this work, each point cloud is divided into \num{32} frames, with each frame containing \num{32} points.

\begin{table*}[]
\caption{{Identification accuracy (${\%} $) across different person and gesture groups in PantoRad and MHomeGes datasets.}}
\label{tabel:user_identification}
\centering
\renewcommand{\arraystretch}{1.1}
\begin{tabular}{@{}ccccccccccc@{}}
\toprule
Dataset & \multicolumn{3}{c}{PantoRad (Single-Hand)} & \multicolumn{3}{c}{PantoRad (Two-Hand)} & \multicolumn{3}{c}{PantoRad (All-Gestures)} & MHomeGes dataset \\ \midrule
Subject Num. & 20   & \ 30   & 40   & \ 20   & \ 30   & 40   & \ 20   & \ 30   & 40   & 25   \\ 
\midrule
Tesla-ID        & 74.4 & \ 57.7 & 50.2 & \ 75.4 & \ 67.2 & 65.0 & \ 72.1 & \ 61.5 & 57.2 & 56.9 \\
Pantomime-ID    & 67.4 & \ 60.0 & 58.9 & \ 66.7 & \ 63.0 & 58.8 & \ 67.0 & \ 58.9 & 52.2 & 80.5 \\
PointNet++-ID   & 91.4 & \ 88.7 & 87.3 & \ 92.7 & \ 91.6 & 89.7 & \ 89.9 & \ 89.5 & 84.4 & 89.1 \\ \bottomrule
\end{tabular}
\end{table*}
Our implementation utilizes PyTorch and PyTorch Geometric, running on a system powered by an NVIDIA GeForce RTX \num{4080} GPU. To mitigate overfitting, an early stopping strategy is employed, configured with a patience threshold of \num{100} epochs. The model optimization is conducted using the Adam optimizer, complemented by a step-decay learning rate schedule, described by the following formulation
{
\begin{equation}
\eta_e = \eta_0 \times \lambda^{\left\lfloor e/T \right\rfloor}\nonumber,
\end{equation}
where \(\eta_e\) is the learning rate at epoch \(e\), \(\eta_0\) represents the initial learning rate (set to \num{0.001}), \(\lambda\) is the decay rate (\num{0.5}), \(T\) indicates the number of epochs after which the learning rate decays (\num{20} epochs), and \(\lfloor \cdot \rfloor\) denotes the floor operation.}
During Autoencoder training, we use the Tesla-GR model as the pre-trained gesture recognition classifier \( G \) and the modified Tesla-ID model as the pre-trained identification classifier \( U \) to generate the {gesture recognition loss} and {identification loss} to train the Autoencoder.


\begin{table*}[t]
\caption{Gesture recognition performance of original point clouds and De-identified point clouds across pantoRad and mHomeGes datasets using accuracy $(\%)$, $F_1$-score, and AUC.}
\label{tab:gr_singlecol}
\centering
\renewcommand{\arraystretch}{1.05}
\begin{tabular}{ccccc ccc ccc}
\toprule
& & \multicolumn{3}{c}{Tesla-GR} & \multicolumn{3}{c}{Pantomime-GR} & \multicolumn{3}{c}{PointNet++-GR} \\
\cmidrule(lr){3-5}\cmidrule(lr){6-8}\cmidrule(lr){9-11}
Dataset & Point Clouds  & Acc.(\%) & $F_1$ & AUC & Acc.(\%) & $F_1$ & AUC & Acc.(\%) & $F_1$ & AUC \\
\midrule
\multirow{2}{*}{PantoRad} & Original Point Clouds & 96.5 & 0.965 & 0.998 & 84.5 & 0.845 & 0.995 & 97.5 & 0.975 & 0.999 \\
                          & De-identified Point Clouds & 87.2 & 0.872 & 0.989 & 58.1 & 0.586 & 0.960 & 74.1 & 0.774 & 0.979 \\
                          \midrule
\multirow{2}{*}{MHomeGes} & Original Point Clouds & 77.3 & 0.772 & 0.969 & 85.7 & 0.857 & 0.987 & 93.6 & 0.937 & 0.996 \\
                          & De-identified Point Clouds & 71.4   & 0.713 & 0.956 & 78.8 & 0.783 & 0.976 & 86.8 & 0.868 & 0.987 \\
\bottomrule
\end{tabular}
\end{table*}

\begin{table*}[t]
\caption{Identification performance of original point clouds and de-identified point clouds on pantoRad and mHomeGes datasets using accuracy $(\%)$, $F_1$-score, and AUC.}
\label{tab:id_singlecol}
\centering
\renewcommand{\arraystretch}{1.05}
\begin{tabular}{ccccc ccc ccc}
\toprule
& & \multicolumn{3}{c}{Tesla-ID} & \multicolumn{3}{c}{Pantomime-ID} & \multicolumn{3}{c}{PointNet++-ID} \\
\cmidrule(lr){3-5}\cmidrule(lr){6-8}\cmidrule(lr){9-11}
Dataset & Point Clouds  & Acc.(\%) & $F_1$ & AUC & Acc.(\%) & $F_1$ & AUC & Acc.(\%) & $F_1$ & AUC \\
\midrule
\multirow{2}{*}{PantoRad} & Original Point Clouds & 57.2 & 0.569 & 0.971 & 52.2 & 0.512 & 0.965 & 84.4 & 0.841 & 0.996 \\
                          & De-identified Point Clouds &  4.1 & 0.019 & 0.758 & 19.3 & 0.151 & 0.840 & 24.9 & 0.217 & 0.868 \\
                          \midrule
\multirow{2}{*}{MHomeGes} & Original Point Clouds & 52.5 & 0.460 & 0.963 & 80.5 & 0.806 & 0.991 & 89.1 & 0.891 & 0.993 \\
                          & De-identified Point Clouds &  13.5    & 0.114 & 0.813 & 34.5 & 0.333 & 0.897 & 53.2 & 0.522 & 0.930 \\
\bottomrule
\end{tabular}
\end{table*}

\section{Results and Analysis} \label{section6}
We evaluate the feasibility and performance of identification using point-cloud gesture data, as well as the effectiveness of ImmCOGNITO in preserving gesture recognition accuracy and decreasing identification performance.

\subsection{Identification Results}\label{section:UI}
Table~\ref{tabel:user_identification} presents the identification accuracy of the three classifiers: Tesla-ID, Pantomime-ID, and PointNet++-ID within the PantoRad and MHomeGes datasets.
For the PantoRad dataset, experiments are conducted with \numlist{20;30;40} subjects in three subsets: Single-Hand, Two-Hand, All-Gestures.
Among the classifiers, PointNet++-ID achieves the highest accuracy, particularly on the Two-Hand subset, with up to \qty{91.4}{\percent} for the \num{20} subjects group.
The Two-Hand gestures consistently yield slightly higher accuracy than the Single-Hand gestures for all identifiers.
The reason is that the Two-Hand gestures contain richer and more person-specific features, offering a more informative dataset to train robust identification models.


As the number of subjects increases, identification accuracy declines across all gesture groups and classifiers.
This decline occurs because a larger pool introduces greater diversity in individual characteristics, making it more challenging for classifiers to distinguish between subjects.
The decline is most pronounced for Single-Hand gestures, as their limited distinctive features provide insufficient information for differentiation.
Tesla-ID's accuracy drops from \qty{74.4}{\percent} (\num{21} subjects) to \qty{50.2}{\percent} (\num{41} subjects).
In contrast, the Two-Hand gestures are less affected by the growing pool of subjects.
For instance, the accuracy of PointNet++-ID drops from \qty{92.7}{\percent} (\num{21} subjects) to \qty{89.7}{\percent} (\num{41} subjects).

For the MHomeGes dataset, which contains \num{25} subjects, the identification accuracy follows a similar trend. PointNet++-ID achieves the highest accuracy (\qty{82.8}{\percent}), while Tesla-ID shows the lowest accuracy (\qty{52.3}{\percent}). Notably, Pantomime-ID achieves \qty{66.8}{\percent}, showing competitive performance in this dataset. The results demonstrate that identification through gesture data is feasible, achieving classification accuracies significantly above the level of random chance. 

\begin{figure*}
    \centering
        \subfloat[Original point cloud.]{\includegraphics[width=0.4\textwidth]{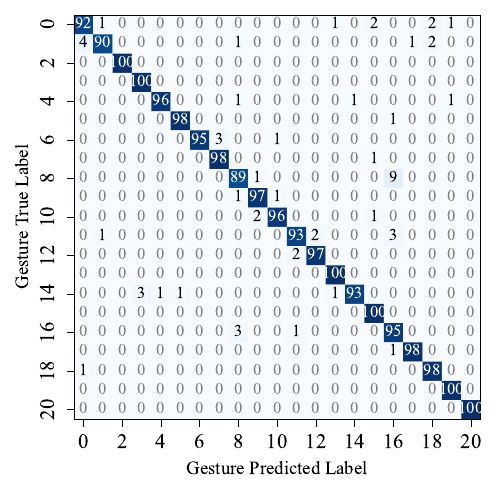}
        \label{fig:cm_for_GR_opc}}
        \hspace{0.05\textwidth}
        \subfloat[De-identified point cloud.]
{\includegraphics[width=0.4\textwidth]{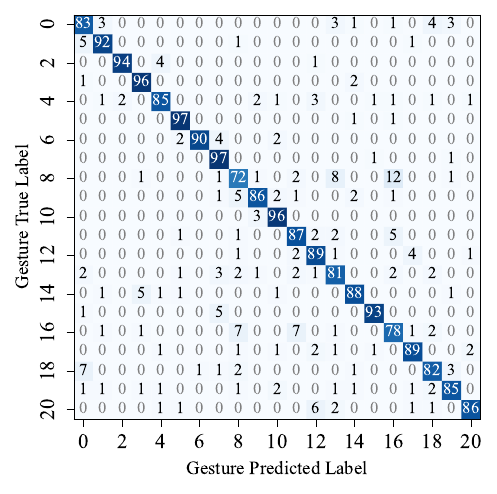}
        \label{fig:cm_for_GR_dpc}}
    \caption{Confusion matrices of the Tesla-GR model for gesture recognition on the PantoRad dataset.}
    \label{fig:cm_for_GR}
\end{figure*}

\begin{figure*}
    \centering
        \subfloat[Original point cloud.]{\includegraphics[width=0.4\textwidth]{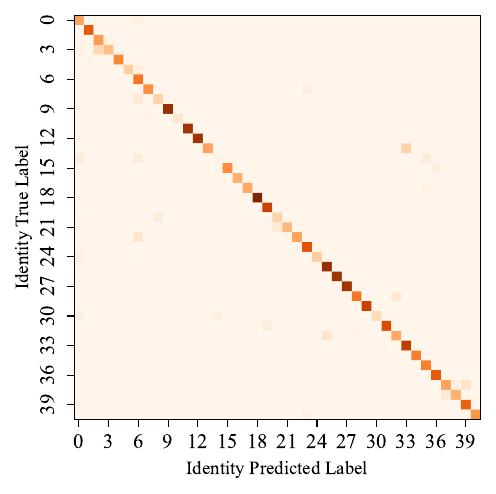}
        \label{fig:cm_for_ui_opc}}
        \hspace{0.05\textwidth}
        \subfloat[De-identified point cloud.]{\includegraphics[width=0.4\textwidth]{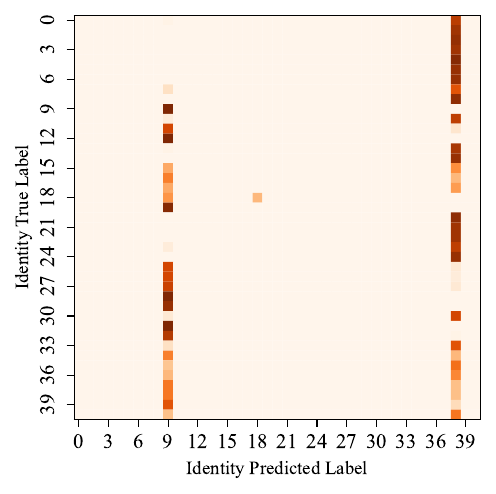}
        \label{fig:cm_for_ui_dpc}}
    \caption{Confusion matrices of the Tesla-ID model for identification on the PantoRad dataset.}
    \label{fig:cm_for_ui}
\end{figure*}

\subsection{Overall Performance using ImmCOGNITO}
{Tables~\ref{tab:gr_singlecol} and \ref{tab:id_singlecol} summarize performance on PantoRad and MHomeGes Datasets. 
Specifically, it compares the performance of both original point clouds and de-identified point clouds in gesture recognition and identification tasks.
The evaluation is conducted using multiple metrics, including Accuracy ($\%$), $F_1$-Score, and AUC, to assess the ability of ImmCOGNITO to maintain accurate gesture recognition while ensuring effective de-identification.

For the PantoRad dataset, {Tesla-GR} on original point clouds exhibits excellent performance in gesture recognition, achieving Accuracy, $F_1$-Score, and AUC of \qty{96.5}{\percent}, \num{0.965}, and \num{0.998}, respectively.
Following the application of the de-identification process, these metrics experienced a slight decline but still maintained high effectiveness, with Accuracy at \qty{87.2}{\percent} and an AUC of \num{0.989}.  
A similar preservation is observed for {Pantomime-GR} and {PointNet++-GR}. These results indicate that ImmCOGNITO largely preserves gesture recognition.
For identification, ImmCOGNITO substantially suppresses identification accuracy. 
Tesla-ID decreases from \qty{57.2}{\percent} on original point clouds to \qty{4.1}{\percent} on de-identified point clouds; 
Pantomime-ID decreases from \qty{52.2}{\percent} to \qty{19.3}{\percent}; 
and PointNet++-ID decreases from \qty{84.4}{\percent} to \qty{24.9}{\percent}.


For the MHomeGes dataset, all models exhibit high gesture recognition accuracies in their original point clouds.
Although accuracy decreases following the de-identification process, it remains at acceptable levels.
When examining identification, the Tesla-ID model experienced a significant reduction in accuracy, decreasing from \qty{56.9}{\percent} to \qty{6.3}{\percent}.
A similar impact was observed for both the PointNet++-ID and Pantomime-ID models.

Overall, across both datasets, de-identified point clouds maintain high gesture recognition performance while driving identification accuracy toward low levels, evidencing effective de-identification with bounded utility loss.}


Fig.~\ref{fig:cm_for_GR} illustrates the confusion matrices for gesture recognition using the Tesla-GR classifier on the PantoRad dataset, presenting results for both original point clouds (Fig.~\ref{fig:cm_for_GR_opc}) and de-identified point clouds (Fig.~\ref{fig:cm_for_GR_dpc}).
As predicted in Fig.~\ref{fig:cm_for_GR_opc}, excellent classification accuracy is achieved, with all \num{21} gesture classes exceeding \qty{95}{\percent}.
Fig.~\ref{fig:cm_for_GR_dpc} reveals that the model continues to exhibit robust gesture recognition performance on de-identified point clouds, despite a slight reduction in accuracy across gesture classes compared with Fig.~\ref{fig:cm_for_GR_opc}.
The most significant accuracy decrease is observed for the gesture with index \num{19}, where the accuracy drops markedly from \qty{100}{\percent} to \qty{74}{\percent}.
This discrepancy is likely due to the varying degrees of person-specific features embedded in different gestures, leading to varying impacts of the de-identification process on gesture recognition performance.

Fig.~\ref{fig:cm_for_ui} presents the confusion matrices for identification using the Tesla-ID classifier on the PantoRad dataset, including both original point clouds (Fig.~\ref{fig:cm_for_ui_opc}) and de-identified point clouds (Fig.~\ref{fig:cm_for_ui_dpc}).
The classifier achieves high identification accuracy with original point clouds (cf. Fig.~\ref{fig:cm_for_ui_opc}), as predictions are predominantly concentrated along the diagonal of the confusion matrix.
In contrast, Fig.~\ref{fig:cm_for_ui_dpc} presents the identification results for de-identified point clouds, where a substantial drop in identification accuracy is observed across all persons.
This indicates that identity features have been effectively obfuscated, making it difficult for the classifier to distinguish between different persons. It is noteworthy that in Fig.~\ref{fig:cm_for_ui_dpc}, Subject 18 still attains a relatively higher identification accuracy of \qty{40.5}{\percent}, compared to \qty{100}{\percent} in Fig.~\ref{fig:cm_for_ui_opc}. This suggests that Subject 18 exhibits more distinctive identity-specific characteristics; nevertheless, the substantial reduction demonstrates the effectiveness of the proposed de-identification Autoencoder in suppressing identity information.



\begin{figure*}
    \centering
        \subfloat[Impact of nearest neighbors $k$ (PantoRad dataset)]{\includegraphics[width=0.4\textwidth]{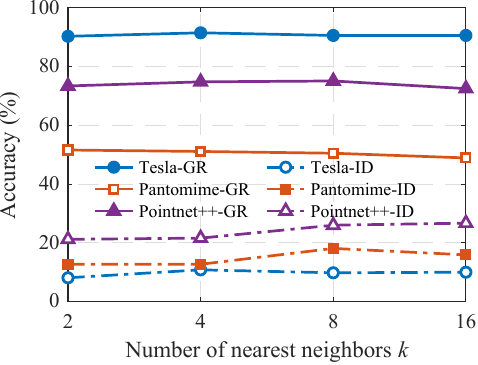}
        \label{fig:loss_factor_1}}
        \hspace{0.05\textwidth}
        \subfloat[Impact of nearest neighbors $k$ (MHomeGes dataset)]{\includegraphics[width=0.4\textwidth]{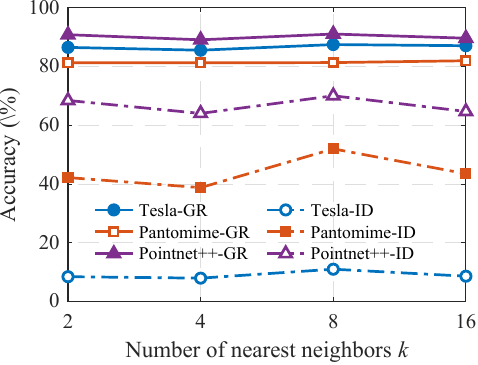}
        \label{fig:loss_factor_2}}
     \caption{Performance of De-identified Point Clouds Across Two Datasets Under Different Numbers of Nearest Neighbors.} 
    \label{fig:neighbors}
\end{figure*}
\begin{figure*}
    \centering
        \subfloat[Impact of loss factor $\beta$ (PantoRad dataset)]{\includegraphics[width=0.4\textwidth]{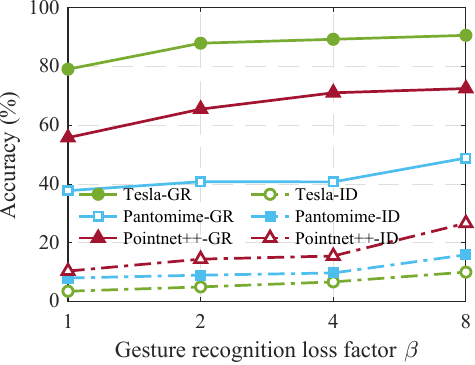}
        \label{fig:loss_factor_3}}
        \hspace{0.05\textwidth}
        \subfloat[Impact of loss factor $\beta$ (MHomeGes dataset)]{\includegraphics[width=0.4\textwidth]{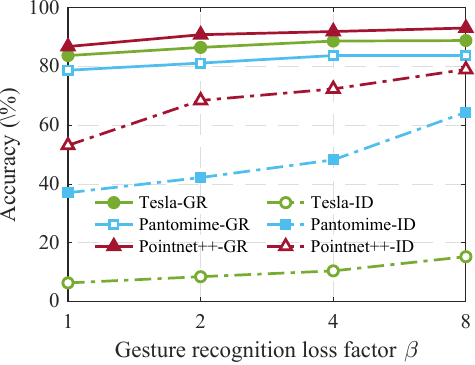}
        \label{fig:loss_factor_4}}
    \caption{Performance of De-identified Point Clouds Across Two Datasets Under Different Gesture Recognition Loss Factors. }
    \label{fig:loss_factor}
\end{figure*}

{
\subsection{Comparison with Perturbation-Based and Established Privacy-Preserving Methods}
To further evaluate the effectiveness of the proposed autoencoder, we compare it with both simple perturbation-based baselines and established privacy-preserving methods on the mmHomeGes dataset. The simple perturbation baselines include additive noise and geometric perturbations applied directly to point clouds, while the established methods include Laplacian noise and k-anonymity. As a reference, we also report performance on the original point clouds without any privacy processing.

\begin{table}[t]

\begingroup
\caption{Comparison with perturbation-based and established privacy-preserving methods on the mmHomeGes dataset, reporting gesture recognition and identification accuracy $(\%)$ evaluated using Tesla-GR and Tesla-ID.}
\label{tab:perturbation}
\centering
\renewcommand{\arraystretch}{1.05}
\setlength{\tabcolsep}{3pt}
\resizebox{\linewidth}{!}{
\begin{tabular}{lcc}
\toprule
 & Gesture Recognition (\%) & Identification (\%) \\
\midrule
Original Point Clouds        & 77.3 & 52.5 \\
\midrule

\multicolumn{3}{l}{\textbf{Point-level Perturbations}} \\
Gaussian Noise        & 67.7 & 47.3 \\
Uniform Noise         & 75.4 & 51.7 \\
Random Perturbation   & 73.7 & 51.3 \\
Scale Perturbation    & 76.5 & 52.2 \\
Rotation Perturbation & 76.2 & 51.5 \\
\midrule
\multicolumn{3}{l}{\textbf{Privacy-oriented Transformations}} \\
Feature Obfuscation   & 67.2 & 47.0 \\
Quantization          & 66.9 & 50.4 \\
\midrule
\multicolumn{3}{l}{\textbf{Strong Privacy / Anonymization}} \\
Laplacian Noise       & 12.7 & 27.8 \\
K-Anonymity           & 11.4 & 7.4  \\
\midrule
\multicolumn{3}{l}{\textbf{Our Method}} \\
ImmCOGNITO (Ours)     & 71.4 & 13.5 \\
\bottomrule
\end{tabular}}
\endgroup
\end{table}

Gesture recognition and identification accuracy are evaluated using Tesla-GR and Tesla-ID, respectively. The results are summarized in Table~\ref{tab:perturbation}. Overall, point-level geometric perturbations, including additive noise and global transformations, largely preserve gesture recognition performance (e.g., 73.7\% under random perturbation and 76.5\% under scaling), but offer limited identity suppression, with identification accuracy remaining close to that of the original point clouds (around 47--52\%).
Feature-level transformations such as feature obfuscation and quantization show a similar trend, where gesture recognition accuracy remains above 66\% while identification accuracy is still high, indicating insufficient de-identification. In contrast, stronger privacy-oriented mechanisms can substantially reduce identification accuracy, but often at the cost of severely degrading gesture recognition performance. For example, Laplacian noise and k-anonymity reduce gesture recognition accuracy to around 12\%.
Compared to these baselines, ImmCOGNITO achieves a more favorable privacy--utility trade-off by reducing identification accuracy from 52.5\% to 13.5\% while maintaining gesture recognition accuracy at 71.4\%.}

\subsection{Impact of the Number of Nearest Neighbors}
In ImmCOGNITO, the parameter \(k\) specifies the number of nearest neighbors for each point with respect to the previous frame in the Temporal Graph \gls{knn}.
This parameter is essential for establishing the local connectivity that captures the critical spatio-temporal dynamics necessary for accurate gesture data interpretation.
Fig.~\ref{fig:neighbors} illustrates the performance of the Autoencoder on the PantoRad dataset.
For this evaluation, point cloud data is generated using various \(k\) values (\numlist{2;4;8;16}) and evaluated across three gesture recognition and identification algorithms.
Fig.~\ref{fig:neighbors} shows that, although an increasing \(k\) enhances the spatio-temporal connections to some extent, it does not significantly improve the overall system performance.
Importantly, as \(k\) increases, the number of edges in the graph constructed by the Temporal Graph \gls{knn} algorithm also increases, thereby raising the computational complexity.
Therefore, we select \(k=2\) in this study.

\subsection{Impact of the Loss Factors}
To achieve a balanced optimization between point cloud reconstruction, gesture recognition preservation, and de-identification, we used a training loss function that integrates point cloud reconstruction loss, gesture recognition loss, and de-identification loss (cf. Section~\ref{section4}).
In ImmCOGNITO, $\alpha$, $\beta$, and $\gamma$ represent the weight coefficients for these three loss components, respectively.
By fixing $\alpha = 1$ and $\gamma = 1$, we systematically adjusted $\beta$ at values of \numlist{1;2;4;8}, analyzing its impact on gesture recognition accuracy and de-identification effectiveness.

Fig.~\ref{fig:loss_factor} presents the performance of the Autoencoder’s de-identified point clouds on the PantoRad and MHomeGes datasets.
As $\beta$ increases, gesture recognition accuracy improves across both datasets, indicating that by enhancing the weight of the gesture recognition loss, more accurate gesture identification can be achieved.
However, increasing $\beta$ also potentially impacts the effectiveness of de-identification.
Enhancing gesture recognition performance concurrently weakens the capability of protecting privacy.

Consequently, ImmCOGNITO can be adapted to various requirements.
For privacy-sensitive scenarios, reducing $\beta$ helps to strengthen de-identification performance.
Conversely, for applications where accurate gesture recognition is critical, increasing $\beta$ boosts recognition accuracy, albeit at the cost of reduced privacy protection.

\begin{table*}[t]
\begingroup

\caption{Gesture recognition and Identification performance of original point clouds and de-identified point clouds across four different indoor environments using accuracy $(\%)$, $F_1$-Score, and AUC.}
\label{tab:gr_singlecol}
\centering
\renewcommand{\arraystretch}{1.05}
\setlength{\tabcolsep}{3pt}
\makebox[\textwidth][c]{%
\begin{tabular}{lcccclccclccclccc}
\toprule
   & \multicolumn{1}{l}{}       & \multicolumn{3}{c}{Industrial} &  & \multicolumn{3}{c}{Restaurant} &  & \multicolumn{3}{c}{Office} &  & \multicolumn{3}{c}{Open} \\ \cmidrule{3-5} \cmidrule{7-9} \cmidrule{11-13} \cmidrule{15-17} 
         &  Point Clouds      & Acc.(\%)   & $F_1$   & AUC     &  & Acc.(\%)   & $F_1$   & AUC    &  & Acc.(\%)  & $F_1$  & AUC   &  & Acc.(\%) & $F_1$ & AUC   \\ \midrule
\multicolumn{1}{c}{\multirow{2}{*}{\begin{tabular}[c]{@{}c@{}}Gesture \\ Recognition\end{tabular}}} & Original      & 89.8       & 0.899   & 0.997   &  & 91.4       & 0.914   & 0.995  &  & 94.6      & 0.946  & 0.998 &  & 96.7     & 0.966 & 0.998 \\
\multicolumn{1}{c}{}                                                                                & De-identified  & 80.3       & 0.797   & 0.987   &  & 80.2       & 0.796   & 0.985  &  & 88.2      & 0.883  & 0.994 &  & 77.1     & 0.760 & 0.983 \\ 
\midrule
\multirow{2}{*}{Identification}                                                                     & Original      & 83.2       & 0.831   & 0.971   &  & 84.2       & 0.842   & 0.965  &  & 74.5      & 0.742  & 0.986 &  & 57.6     & 0.553 & 0.937 \\
 & De-identified & 28.0       & 0.181   & 0.483   &  & 32.6       & 0.223   & 0.808  &  & 5.1       & 0.020  & 0.676 &  & 10.4     & 0.029 & 0.501 \\
 \bottomrule
\end{tabular}}
\endgroup
\end{table*}

\begin{table*}[t]
\begingroup

\caption{Gesture recognition and identification performance on distance-based subsets of the mmHomeGes dataset, evaluated using Tesla-GR and Tesla-ID in terms of accuracy $(\%)$, $F_1$-Score, and AUC.}
\label{tab:distance}
\centering
\renewcommand{\arraystretch}{1.05}
\setlength{\tabcolsep}{3pt}
\makebox[\textwidth][c]{%
\begin{tabular}{ccccclccclccc}
\toprule
\multicolumn{1}{l}{} & \multicolumn{1}{l}{} & \multicolumn{3}{c}{$S_1$(1.20--2.10)} & \multicolumn{1}{c}{} & \multicolumn{3}{c}{$S_2$(2.25--3.00)} &  & \multicolumn{3}{c}{All(1.20--3.00)} \\ \cmidrule{3-5} \cmidrule{7-9} \cmidrule{11-13} 
\multicolumn{1}{l}{}                                                            & Point Clouds         & Acc.(\%)                & $F_1$                & AUC                  &                      & Acc.(\%)                & $F_1$                & AUC                  &  & Acc.(\%)                & $F_1$               & AUC                 \\ \midrule 
\multirow{2}{*}{\begin{tabular}[c]{@{}c@{}}Gesture \\ Recognition\end{tabular}} & Original             & 81.9                    & 0.818                & 0.973                &                      & 86.3                    & 0.863                & 0.985                &  & 77.3                    & 0.772               & 0.969               \\
& De-identified        & 78.4                    & 0.783                & 0.971                &                      & 82.3                    & 0.824                & 0.980                &  & 71.4                    & 0.713               & 0.956               \\ \midrule
\multirow{2}{*}{Identification}                                                 & Original             & 56.38                   & 0.506                & 0.963                &                      & 56.5                    & 0.5395               & 0.9650               &  & 52.5                    & 0.460               & 0.963               \\ & De-identified        & 33.62                   & 0.334                & 0.760                &                      & 22.6                    & 0.2197               & 0.8078               &  & 13.5                    & 0.114               & 0.813\\       
\bottomrule
\end{tabular}}
\endgroup
\end{table*}

{
\subsection{Impact of the Indoor Environments}
To assess the robustness of the proposed method under varying environmental complexities and multipath conditions, we conducted experiments across four distinct scenarios from the PantoRad dataset, which are {Open}, {Office}, {Restaurant}, and Factory [1].  These environments represent a diverse range of spatial characteristics: the {Open} scenario is a spacious 91 $m^{2}$ area devoid of furniture; the {Office} represents a compact 18 $m^{2}$ workspace with high clutter from nearby objects; the {Restaurant} (23.36 $m^{2}$) introduces tables, chairs, and walls within a 3 m radius; and the {Factory} simulates an industrial setting containing robots and machinery within the same range.

The performance of gesture recognition and identification across these environments is summarized in Table~\ref{tab:gr_singlecol}. The results demonstrate that the proposed autoencoder maintains stability, preserving high gesture recognition performance across all tested conditions. Conversely, identification accuracy is substantially reduced on the de-identified point clouds. It is observed that the {Factory} and {Restaurant} scenarios exhibit slightly higher identification accuracy on de-identified data (around 30\%) compared to the {Office} and {Open} environments. However, this is primarily attributed to the limited subject pool in these subsets (four subjects), which raises the random guessing baseline to 25\%. In addition, the gesture recognition accuracy on de-identified data remains approximately 80\% across all four environments, suggesting that the proposed method retains reasonable performance under different indoor environments.}

{
\subsection{Impact of the Distance}
To investigate the impact of distance on the proposed autoencoder, we conduct additional experiments on the mmHomeGes dataset by evaluating its performance under different distance conditions. The mmHomeGes dataset contains samples collected at distances ranging from 1.2~m to 3.0~m with an interval of 0.15~m. Based on this, we construct three distance-based subsets: $S_1$ (1.20--2.10~m), $S_2$ (2.25--3.00~m), and the full set covering all distances (1.20--3.00~m). For each subset, the autoencoder is trained and evaluated independently. Gesture recognition and identification performance are then assessed using Tesla-GR and Tesla-ID, respectively.

The results are summarized in Table~\ref{tab:distance}. As shown in the table, the proposed autoencoder maintains stable gesture recognition performance across different distance conditions after de-identification. For instance, in subset $S_2$ (2.25--3.00~m), gesture recognition accuracy remains above 82\%, indicating robustness at distances.
In contrast, identification performance is substantially reduced across distance conditions. When aggregating samples from all distances, identification accuracy drops markedly from 51.4\% to 13.5\%, demonstrating that the proposed autoencoder effectively suppresses identity-related information while remaining robust to distance variations.}

{
\subsection{Ablation Studies}
To better understand the contribution of each module in the proposed framework, we conduct extensive ablation studies by systematically removing or replacing individual components. All ablation models are trained using the same hyperparameters and training protocol as the full model to ensure a fair comparison. Specifically, we design the following ablation settings:
\begin{itemize}
    \item \textbf{w/o de-identification Loss}: The identity-related loss is removed by setting $\beta = 0$, to evaluate its contribution to privacy protection.
    
    
    
    \item \textbf{w/o Temporal Graph Edges}: Cross-frame temporal edges are removed, and only spatial graphs within each frame are used to evaluate the importance of temporal modeling.
    
    \item \textbf{w/o Temporal KNN}: Temporal information is excluded from KNN neighbor search, and neighbors are selected based only on spatial coordinates to assess the role of temporal cues in graph construction.
    
    \item \textbf{w/o Max Pooling}: The global max-pooling and feature concatenation module is removed, and the output point cloud is generated using only point-wise features, to evaluate the contribution of global context information.
\end{itemize}

The ablation results are summarized in Table~\ref{tab:ablation}, where gesture recognition and identification performance are evaluated using Tesla-GR and Tesla-ID on de-identified point clouds. Removing the identity loss leads to a clear increase in identification accuracy (from 4.1\% to 45.8\%) while keeping gesture recognition accuracy high (89.7\%), confirming the essential role of the identity loss in de-identification. Eliminating temporal graph edges causes a substantial drop in gesture recognition accuracy from 87.2\% to 71.5\%, highlighting the importance of temporal message passing for capturing gesture dynamics. When temporal information is removed from the KNN construction, gesture recognition accuracy remains high (87.1\%), while identification accuracy increases compared to the full model, indicating that temporal information in KNN helps suppress identity-related cues and improves de-identification.
 Removing the global max-pooling operation slightly degrades performance (gesture recognition accuracy drops to 84.7\%), suggesting that global feature aggregation helps preserve gesture structure. 
Overall, the full model achieves the best privacy--utility trade-off, maintaining competitive gesture recognition accuracy while minimizing identification performance.

\begin{table}[t]
\begingroup

\caption{Ablation study results on the PantoRad dataset, reporting gesture recognition and identification performance evaluated using Tesla-GR and Tesla-ID on de-identified point clouds in terms of accuracy $(\%)$, $F_1$-Score, and AUC.}
\label{tab:ablation}
\centering
\renewcommand{\arraystretch}{1.05}
\setlength{\tabcolsep}{3pt}
\resizebox{\linewidth}{!}{
\begin{tabular}{cccccccc}
\toprule
                         & \multicolumn{3}{c}{Gesture Recognition} &  & \multicolumn{3}{c}{Identification} \\ \cmidrule{1-4} \cmidrule{6-8} 
                         & Acc.(\%)       & $F_1$        & AUC        &  & Acc.(\%)     & $F_1$       & AUC      \\ \midrule
w/o De-identification Loss        & \textbf{89.7}          & 0.896      & 0.995      &  & 45.8        & 0.451     & 0.945    \\
w/o Temporal Graph Edges & 71.5          & 0.705      & 0.961      &  & 6.9         & 0.054     & 0.749    \\
w/o Temporal KNN         & 87.1          & 0.870      & 0.993      &  & 13.1        & 0.109     & 0.738    \\
w/o Max Pooling          & 84.7          & 0.846      & 0.992      &  & 9.9         & 0.072     & 0.811    \\
Full Model                & \textbf{87.2}          & 0.872      & 0.989      &  & \textbf{4.1}         & 0.019     & 0.758   \\
\bottomrule
\end{tabular}}
\endgroup
\end{table}}

\section{Discussion} \label{section7}


\subsection{Application Scenarios, Ethical and Social Impact}

Point cloud data is applicable across IoT settings, including smart homes and healthcare. For example, touchless home control can use swipe/rotate gestures for lighting or media, and de-identified sequences in clinics or elder care can track prescribed exercises or flag abnormal motions without storing stable biometric traits. These sectors often demand stringent privacy protections. The privacy preservation techniques proposed in this study safeguard identity and can be adapted to suit broader applications, enhancing system utility and trust.


\subsection{Limitations and Outlook}
\noindent{\textbf{Identity Obfuscation in Diverse Applications:}}
This study is confined to identity suppression in \gls{mmwave} radar gesture recognition. Our experiments and metrics target gesture utility and identity leakage within IoT interaction scenarios, for example, smart-home lighting/media control and in-vehicle infotainment. Nevertheless, mmWave radar supports many other applications—such as vital-sign monitoring, occupancy estimation, fall detection, and activity recognition—where privacy requirements likewise arise. Future work may extend identity-obfuscation mechanisms to these settings by defining task-specific utility objectives and evaluation protocols and by adapting architectures or training strategies to balance identity suppression with task performance.

\noindent{\textbf{{Multi-subject Environments:}}
We acknowledge that the single-subject assumption constrains deployment in shared spaces. Real-world scenarios, such as ride-sharing cabins where a driver issues gesture commands while a passenger shifts posture, require concurrent de-identification and recognition for multiple users. Subsequent studies could explore multi-instance learning with permutation-invariant encoders and lightweight association and tracking to maintain per-subject gesture streams while limiting cross-subject leakage.

\noindent{\textbf{Comprehensive Privacy in Radar Systems: }}
Privacy concerns in point cloud data extend beyond mere identity information.
Privacy information that could be exposed includes behavior patterns, health conditions, and living environments.
Future research should consider developing multifaceted privacy-enhancing technologies that can prevent the unintentional disclosure of a broad range of sensitive information.
Introducing machine learning models specifically designed to identify and segregate sensitive information from harmless data can help more effectively apply privacy measures.

{
\noindent\textbf{Stronger Adversarial Settings.}
We will extend the evaluation of the proposed framework to stronger adversarial settings that are commonly considered in privacy-preserving learning. This includes open-set identity recognition, identification models with unseen backbones trained from scratch on de-identified point clouds, as well as cross-dataset transfer scenarios. Moreover, adaptive adversaries that explicitly attempt to exploit residual biometric cues in the transformed representations constitute an important direction for further investigation. Addressing these challenges will require more diverse datasets and carefully designed adversarial evaluation protocols, which are beyond the scope of the present study and are left for future work.
}

\section{Conclusion} \label{section8}
In this work, we aim to reduce identity-related information embedded in mmWave radar point cloud data used for gesture recognition in IoT environments. To this end, we propose a graph-based autoencoder framework that transforms the raw point cloud into a structured representation while preserving gesture-discriminative features. Our method constructs a directed graph from the input point cloud, applies message passing neural networks to extract local node features, and uses multi-head self-attention to model global spatial-temporal dependencies. The encoded representation is then used to reconstruct a modified point cloud with reduced identity information. Experimental results show that point clouds reconstructed by our method significantly lower the performance of identity classifiers while maintaining high gesture recognition accuracy, demonstrating the feasibility of identity information suppression in point cloud data. These results support privacy-preserving gesture interaction in IoT scenarios, such as smart homes and in-vehicle interfaces.



\end{document}